\documentclass[useAMS,usenatbib]{mn2e}
\usepackage{times} 
\usepackage{graphicx}
\usepackage{amsmath,amssymb}
\usepackage{xspace}
\usepackage{framed}
\usepackage{color}
\usepackage{ifthen}
\usepackage{natbib}
\usepackage{multirow}
\usepackage{subfigure}
\usepackage{xspace}
\usepackage{hyperref}

\newcommand{\be}{\begin{equation}}
\newcommand{\ee}{\end{equation}}
\def\ba#1\ea{\begin{align}#1\end{align}}
\newcommand{\msun}{{\rm M}_\odot}
\newcommand{\msunm}{~{\rm M}_\odot~}
\newcommand{\mrange}[3][M]{$#2\,{\rm M}_\odot\leq #1 < #3\,{\rm M}_\odot$}
\newcommand{\mlim}[1]{$M>#1{\rm M}_\odot$}
\newcommand{\mlow}[1]{$M<#1{\rm M}_\odot$}

\def\Te{T_{\rm eff}}
\def\logg{\log g}
\def\masyr{\mbox{$~{\rm mas}\,{\rm yr}^{-1}$}}
\def\kms{\mbox{$~{\rm km}\,{\rm s}^{-1}$}}
\def\kpc{\mbox{$~{\rm kpc}$}}
\def\rsun{\,R_\odot}

\def\bsefr{\Gamma_{\rm BSE}}
\def\ssefr{\Gamma_{\rm SSE}}
\newcommand{\gyr}{\ensuremath{{\rm~Gyr}\xspace}}
\newcommand{\bse}{\texttt{BSE}~}
\newcommand{\seba}{\texttt{SeBa}~}

\renewcommand{\b}[1]{\bmath{#1}}
\makeatletter
\newcommand{\Rmnum}[1]{\expandafter\@slowromancap\romannumeral #1@}
\makeatother
\title[White Dwarf Kinematics vs Mass]{White Dwarf Kinematics vs Mass}
\author[C. Wegg and E. S. Phinney]{Christopher Wegg$^{1}$\thanks{E-mail:
wegg@tapir.caltech.edu} and E. Sterl Phinney$^{1}$ \\
$^{1}$Department of Physics, Mathematics and Astronomy, MC 350-17, California Institute of Technology, Pasadena, CA 91125, USA }

\begin{document}

\pagerange{\pageref{firstpage}--\pageref{lastpage}} \pubyear{2011}
\maketitle
\label{firstpage}

%-----------------------------------------------------------------------
\begin{abstract}
We have investigated the relationship between the kinematics and mass of young ($<3\times10^8$ years) white dwarfs using proper motions. Our sample is taken from the colour selected catalogues of SDSS \citep{Eisenstein:2006} and the Palomar-Green Survey \citep{Liebert:2005}, both of which have spectroscopic temperature and gravity determinations. We find that the dispersion decreases with increasing white dwarf mass. This can be explained as a result of less scattering by objects in the Galactic disk during the shorter lifetime of their more massive progenitors. A direct result of this is that white dwarfs with high mass have a reduced scale height, and hence their local density is enhanced over their less massive counterparts. In addition, we have investigated whether the kinematics of the highest mass white dwarfs ($>0.95\msunm$) are consistent with the expected relative contributions of single star evolution and mergers. We find that the kinematics are consistent with the majority of high-mass white dwarfs being formed through single star evolution. 
\end{abstract}
%-----------------------------------------------------------------------

%-----------------------------------------------------------------------
\section{Introduction}
\label{sec:intro}
%-----------------------------------------------------------------------

Despite the significant work on both the kinematics and mass distribution of white dwarfs, very little work has addressed their connection.

The kinematics of galactic white dwarfs have been studied on numerous occasions with several motivations. They have proven useful in attempts to unravel the evolutionary history and progenitors of the various classes of white dwarfs \citep{Sion:1988,Anselowitz:1999}. Interest in white dwarf kinematics was also prompted by the suggestion that halo white dwarfs could provide a significant contribution to Galactic dark matter \citep{Oppenheimer:2001, Reid:2005}. This effort has concentrated on the identification of halo white dwarfs and estimating the resultant density, which now appears to be a small contribution to the Galactic dark matter budget \citep{Pauli:2006}. Moreover, the mass distribution of the most common hydrogen rich (DA) white dwarfs has also been extensively investigated, particularly for white dwarfs with $T\gtrsim10$,$000$\,K which are hot enough for their masses to be deduced spectroscopically from fits to their Balmer lines \citep{Liebert:2005,Kepler:2007,Vennes:1999}. The mass distribution shows a peak at $0.6 {\rm M}_\odot$ due to the relative abundance of their lower mass progenitors with a tail extending to higher masses formed from more massive progenitors.

The connection between the galactic kinematics of a group of thin disk objects and their progenitors is largely due to the process of kinematic disk `heating' \citep{Wielen:1977,Nordstrom:2004}. The hot white dwarfs with short cooling ages we observe in the galactic neighbourhood today are formed from a wide range of progenitor masses ($\sim0.8$--$8\, \msun$) and hence have a wide range in age. We therefore expect high-mass disk white dwarfs to have a low velocity dispersion in comparison to low-mass disk white dwarfs whose progenitors formed earlier. This connection was suggested in \citet{Guseinov:1983} who performed an analysis suggesting that white dwarfs with larger masses have smaller dispersions, however this was reinvestigated by \citet{Sion:1988} with a larger sample of 78 DA white dwarfs where no evidence for any correlation was found. This paper readdresses the connection between mass and kinematics with a greatly increased sample size.

The outline of the paper is as follows: In section \ref{sec:sample} we discuss the sample selection and the calculation of distances and proper motions. In section \ref{sec:kinomethod} we discuss how we estimate the kinematics of the sample without radial velocity information. We use two methods, that of \citet{Dehnen:1998} (section \ref{sec:dehnen}), and a Markov Chain Monte Carlo (MCMC) where we marginalise over the unknown radial velocity (section \ref{sec:mcmc}). In section \ref{sec:sse} we analyse whether the kinematics are consistent with single star evolution (SSE) both via analytic methods (section \ref{sec:sseanalytic}) and simulations (section \ref{sec:montesse}). In section \ref{sec:bse} we analyse whether the highest mass white dwarfs are largely formed through single star evolution or are the product of the merger of two lower mass white dwarfs. Finally, we discuss the implications of our findings on the scale height of white dwarfs in section \ref{sec:scaleh}.

For the reader in a hurry, the primary result of this paper, the relationship between the mass of young white dwarfs and their velocity dispersion, is shown in figure \ref{fig:results} and discussed in section \ref{sec:kinomethod}. The implied scale heights, the second key result, are then discussed in section \ref{sec:scaleh}. These results have been checked using a Monte Carlo simulation of the formation and observation of an ensemble of white dwarfs, which is described by flowcharts in figures \ref{fig:sseflow}--\ref{fig:obsflow}: in figure \ref{fig:sseflow} the process of choosing stars is described, in figure \ref{fig:diskflow} the process of placing them in the disk is described, and in figure \ref{fig:obsflow} the process of determining the observability of the simulated white dwarf is described.

%-----------------------------------------------------------------------
\section{Sample}
\label{sec:sample}
%-----------------------------------------------------------------------

We investigate only hydrogen atmosphere (DA) white dwarfs due to the relative simplicity of their spectra and the resultant security of the spectroscopic masses. The sample of DA white dwarfs is taken from two sources, the Palomar-Green (PG) white dwarf survey \citep{Liebert:2005} and the SDSS DR4 white dwarf survey \citep{Eisenstein:2006}. The SDSS sample is much larger the the PG sample. The PG sample is included as a demonstration that the results are secure, and not a result of systematics in SDSS, such as the complex selection of targets. For clarity we first discuss which types of white dwarfs we select, then discuss how the SDSS survey is dealt with, and finally how the PG survey was dealt with. The sample and its selection is summarised in table \ref{tab:sample}.

{\sc Selected White Dwarfs:} Both PG and SDSS are colour selected, eliminating the kinematic biases inherent in proper motion based surveys, and contain spectroscopic determinations of surface gravity, $\logg$, and effective temperature, $\Te$, obtained by fitting the profile of the Balmer lines. We restrict the sample to objects whose fitted $\Te$ was between 13,000\,K and 40,000\,K, since $\logg$ appears to be systematically overestimated at low temperatures and $\Te$ overestimated at higher temperatures \citep{Eisenstein:2006}. 

The fitted $\logg$ and $\Te$ are converted to masses and ages using the models of the carbon core white dwarf cooling models of \citet{Fontaine:2001} below 30,000\,K and
 \citet{Wood:1995} with thick hydrogen layers of fractional mass $10^{-4}$ above 30,000\,K  \footnotemark[1]. White dwarfs with inferred masses less than $0.47\,\msun$ are instead assumed to have helium cores whose masses and ages are calculated from the models of \citet{Serenelli:2001}. Only objects with cooling ages below $3\times10^8$~years are included in the sample to avoid significant kinematic heating after white dwarf formation. The requirements of cooling age below $3\times10^8$~years and $\Te$ above 13,000\,K are competing. Above $0.60~\msun$ the WDs cool more slowly and thus the age limit is used, while below $0.60~\msun$ the temperature limit is used.

White dwarfs previously discussed in the literature as known members of binaries were removed from the samples.

{\sc SDSS Survey \citep{Eisenstein:2006}:} Many of the SDSS spectra have low signal-to-noise ratios and hence large errors on their fitted $\logg$ and $\Te$. To ensure accurate masses and photometric distances only objects whose spectra had a signal-to-noise ratio larger than 10 are included. The grid of model atmospheres fitted in the SDSS catalog extends only to $\logg=9$, and thus, for objects at this limit, the refitted $\logg$ and $\Te$ given in \cite{Kepler:2007} were used.

Photometric distances to the white dwarfs in SDSS are calculated by minimising
\begin{align}
\chi^2 = \sum_{i=(u,g,r,i,z)} ( m_i - [&M_i(\logg,\Te) + \nonumber \\ 
&A_g a_i + 5 \log d - 5 ] )^2/\sigma_i^2
\label{eq:chi2}
\end{align}
where $m_i$ and $\sigma_i$ are the 5 band SDSS photometry and their errors, $M_i$ are the model absolute magnitudes, $A_g a_i$ is the reddening and $d$ the distance in parsecs. The photometric $\sigma_i$ is the quoted photometric error in SDSS each band added in quadrature to a systematic error of $(u,g,r,i,z)=(0.015,0.007,0.007,0.007,0.01)$~\citep{Kleinman:2004}. Model absolute magnitudes are taken from the atmospheric models provided by Bergeron\footnote{Available from \url{http://www.astro.umontreal.ca/~bergeron/CoolingModels/}, uses results from from \citet{Holberg:2006}, \citet{Kowalski:06}, \citet{Tremblay:11} and \citet{Bergeron:11}}. $A_g a_i$ is the product of $R_V=3.1$ extinction in each band of $(a_u,a_g,a_r,a_i,a_z)=(1.36, 1.00, 0.73, 0.55, 0.39)$ and the overall extinction $A_g$, which is constrained to lie between zero and the value of galactic extinction map of \citet{Schlegel:1998} at the position of the object considered. 

\begin{figure}
\includegraphics[width=\linewidth]{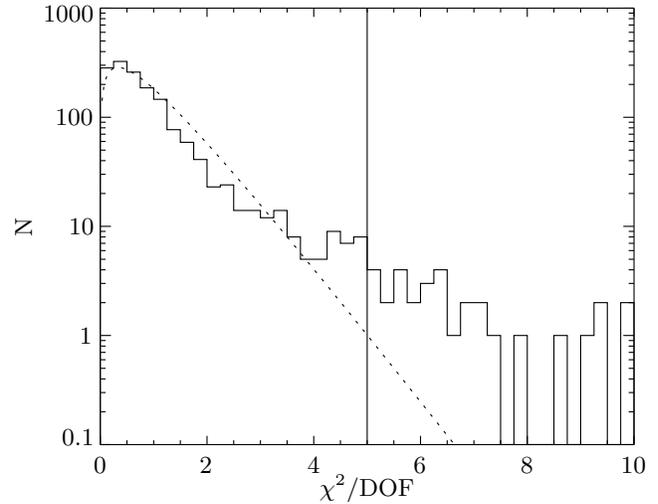}
\caption{$\chi^2$ per degree of freedom (DOF) for the fitted photometric distance of the $1443$ SDSS DA white dwarfs considered. A $\chi^2$ function with $3$ DOF is plotted as the dotted line. Beyond $\chi^2=5$ the white dwarfs are rejected.} 
\label{fig:distch2}
\end{figure}

The resulting distribution of $\chi^2$ values calculated by minimising equation \ref{eq:chi2} is plotted in figure \ref{fig:distch2}. It closely resembles a $\chi^2$ distribution, but with an extended tail. Objects with reduced $\chi^2$ larger than 5 
were removed from the sample, most of these objects show an excess towards the redder photometric bands, indicating they are in binaries with a cooler white dwarf companion. Errors in the photometric distance are taken to be the $\Delta \chi^2 = 1$ surface added in quadrature to the distance errors introduced though the uncertainty in $\logg$ and $\Te$. 

Proper motions for the SDSS sample are taken from the catalogue of \citet{Munn:2008}. These proper motions are calculated from the USNO-B1.0 plate positions re-calibrated using nearby galaxies together with the SDSS position so that the proper motions are more accurate and absolute. By measuring the proper motions of quasars \citet{Munn:2004} estimates that the $1\sigma$ error is 5.6\masyr.

{\sc PG Survey:} For 132 stars in the PG survey, SDSS photometry was available and the same method was used as for SDSS stars. For the remaining objects the PG catalog photometric distances were used. These were estimated in \citet{Liebert:2005} from comparison of the $V$ band magnitude with the predicted $M_V$ from the same models of \citet{Holberg:2006}. Comparison of the stellar distances given by the two methods gives a standard deviation of 7 per cent. The majority of this error is expected to be in the PG survey distances and hence a conservative 10 per cent error was applied to these.

Proper motions for PG white dwarfs that appear in SDSS are taken from the catalog of \citet{Munn:2008}. For the remaining objects, the PPMXL proper motion was used where available, which has typical $1\sigma$ error of $\sim$8\masyr~\citep{Roeser:2010}.

Finally 4 objects in the PG sample have no reliable PPMXL proper motion, primarily due to a spurious matching of objects between epochs. For these, the proper motion was calculated directly between the scanned POSS-\Rmnum{1} and POSS-\Rmnum{2} plates. The proper motion was measured relative to nearby faint stars of similar magnitude corrected for galactic rotation (see Section \ref{sec:dehnen}). 
Typical errors estimated from the proper motions of stars of similar magnitude to be $11$\masyr. We emphasize that only 4 of 1491 white dwarfs use this method, and none have mass above $0.95~\msun$ analyzed in more detail in Section \ref{sec:bse}.

\begin{figure}
\includegraphics[width=\linewidth]{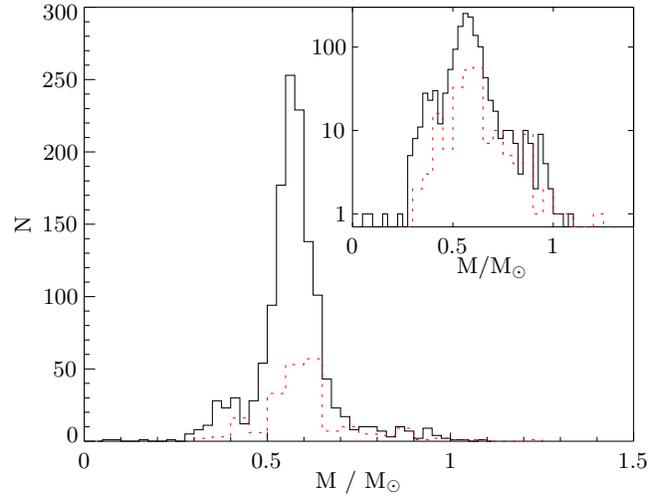}
\caption{Mass distribution of the samples of SDSS (black line) and PG (dashed red line) white dwarfs after the cuts described in the text. Inset graph shows the same data on logarithmic axes.} 
\label{fig:massdist}
\end{figure}

{\sc Final Sample:} The resulting sample of 1443 SDSS  and 211 PG white dwarfs contains young DA white dwarfs with reliable masses, proper motions and photometric distances. The mass distribution of the samples is shown in figure \ref{fig:massdist}. The process of constructing the sample together with numbers of objects is summarised in table \ref{tab:sample}.

\begin{table}
\begin{tabular}[l]{l c c}
\hline
    & PG & SDSS \\
\hline
  Number of DA White Dwarfs with & \multirow{2}{*}{299} & \multirow{2}{*}{6926} \\
  good photometry not known to be binaries &           &          \\
  of these number with signal-to-noise $>10$ &          299 &         3125 \\
  of these number with $13,000K<\Te<40,000K$ &          215 &         1555 \\
  of these number with age $<3\times10^8$ yrs &          211 &         1491 \\
\hline
Distance source: & & \\
  \citet{Liebert:2005} &           79& 0 \\
  SDSS Photometry &          132&         1491 \\
  of these number rejected with $\chi^2>5$ & 0 &           48 \\
\hline
Proper Motion Source: & & \\
  \citet{Munn:2008} &          153 &         1443\\
  PPMXL &           54 & 0\\
  Manual measurement from POSS I/II &            4 & 0\\ 
\hline
\end{tabular}
\caption{Summary of sample}
\label{tab:sample}
\end{table}

%-----------------------------------------------------------------------
\section{Kinematics Without Radial Velocities}
\label{sec:kinomethod}
%-----------------------------------------------------------------------

We now turn to calculating the mean velocity and the velocity dispersion for our sample. While radial velocities are required to completely determine the kinematics of an individual object, bulk kinematic properties such as the mean velocity and the velocity dispersion can be determined from only transverse motions. 

We use two methods to do so, the frequentist method used in section \ref{sec:dehnen}, and a Markov Chain Monte Carlo in section \ref{sec:mcmc}. Both methods give similar results which are summarised in table \ref{tab:results}. 

%-----------------------------------------------------------------------
\subsection{Method of Dehnen and Binney (1998)}
\label{sec:dehnen}
%-----------------------------------------------------------------------

The method used here is adapted from \citet{Dehnen:1998}. First the observed proper motions in galactic coordinates,  $\mu_{\ell}^{\rmn{obs}}$ and $\mu_{b}^{\rmn{obs}}$, are corrected for Galactic rotation through
\begin{equation} \begin{array}{lcl}
 \mu_\ell &=& \mu_{\ell}^{\rmn{obs}} - A \cos(2\ell)  - B \\ 
 \mu_b &=& \mu_{b}^{\rmn{obs}} + A \sin(2\ell)\cos b \sin b.
\label{eq:rotcor}
\end{array} \end{equation}
using  $A=14.82 \kms\kpc^{-1}$ and $B=-12.37\kms\kpc^{-1}$ \citep{Feast:1997}. In galactic coordinates
where the components are directed towards the galactic centre, in the direction of galactic rotation, and towards the north Galactic pole we observe the velocity
\begin{equation} \b{V}_\perp = 4.74 d \left[\begin{array}{c}
	 -\mu_\ell \,\sin\ell\,\cos b - \mu_b \, \cos\ell\,\sin b \\
	  \mu_\ell \, \cos\ell\,\cos b - \mu_b \, \sin\ell\,\sin b \\
					         \mu_b\,\cos b
		   \end{array} \right]\kms
\end{equation}
with $d$ in kpc and  proper motions in\masyr. This is the projection of the velocity $\b{V}$ onto the sky plane though the projection matrix 
\begin{equation}
\b{V}_\perp = \b{A} \cdot \b{V}\,, \quad \b{A}=\b{I}-\hat{\b{r}}\otimes\hat{\b{r}}
\end{equation}
where $\hat{\b{r}}$ is the unit vector to the star. 

Next the quantity $S^2$ is formed through
\begin{equation}
S^2(\b{V_0}) \equiv \left\langle | \b{V}_\perp - \b{A}\cdot \b{V_0} |^2 \right\rangle .
\label{eq:S2}
\end{equation} 
Under the assumption that the positions of the observed objects are uncorrelated with the velocity, then the choice of $\b{V_0}$ that minimises $S^2$ is the mean velocity. Also $S^2$ at the minimum is a measure of the dispersion of the group.

\citet{Dehnen:1998} then calculate all independent six elements of the dispersion tensor. Unfortunately, this entails estimating nine parameters which limits its use to samples with large numbers of objects. This would require excessively wide bins for the high-mass region where there are few objects. Instead we choose to make further assumptions about the objects' velocities in order to reduce the number of fitted parameters. The mean velocity of each group towards the galactic centre and the north Galactic pole is simply a result of the solar motion and we take these to be 10.00\kms~ and 7.17\kms~ respectively \citep{Dehnen:1998}. The mean velocity in the direction of galactic rotation, $V_0$, is kept as a free parameter since in addition to the solar motion this varies between groups due to asymmetric drift. We also assume that the dispersion tensor takes the form 
\begin{equation}
\b{\sigma} = \sigma_1 \mathrm{diag} \left(1, \frac{1}{1.4}, \frac{1}{2.2}\right) 
\end{equation}
which is accurate for main sequence stars in the solar neighbourhood \citep{Dehnen:1998}. This reduces the number of parameters for each group to the asymmetric drift $V_0$ and the normalisation of the dispersion tensor $\sigma_1$. 

$V_0$ is calculated by minimising equation \ref{eq:S2}, and then $\sigma_1$ is estimated though a Monte-Carlo simulation: Since $S^2$ is a measure of the dispersion, an initial estimate of $\sigma_1^2$ is taken to be $S^2$, and a set of simulations is performed where a new velocity is chosen for each white dwarf at its position in the sky from the isothermal distribution with the assumed dispersion tensor and the calculated mean velocity. The error in tangential velocity, assumed to be Gaussian, is added to this. The set of simulations produces a distribution of $S^2$ values, and $\sigma_1^2$ is iterated until the mean $S^2$ corresponds to the value calculated from observations. $S^2$ is almost proportional to $\sigma_1^2$ when errors in tangential velocity are neglected and so the error in $\sigma_1^2$ is estimated from the distribution of $S^2$ scaled by this proportionality constant.

%-----------------------------------------------------------------------
\subsection{MCMC Estimate}
\label{sec:mcmc}
%-----------------------------------------------------------------------
 
In addition, a Markov Chain Monte Carlo (MCMC) likelihood based estimate of the kinematic parameters was obtained. We use uninformative flat priors for the fitted parameters.

We denote the probability that the velocity of the $i$th object was $\b{V}$ to be $P(\b{V} | \b{D}_{i},\b{\sigma}_i)$ where $\b{D}_{i}=(l,b,d,\mu_\ell, \mu_\rmn{b})$ is the data for the $i$th object together with the corresponding errors $\b{\sigma}_i$. $\mu_\ell$ and $\mu_\rmn{b}$ are the values corrected for galactic rotation by equation \ref{eq:rotcor}. Under the assumption that positions are uncorrelated with velocity then the distribution function is a function only of velocity: $f(\b{V})$. In addition, in what follows we do not consider the positions, but instead focus on the kinematics through the velocity $\b{V}$. Under these assumptions the overall likelihood for a set of observations of a group of white dwarfs is
\begin{align}
{\cal L}\, &= \, \prod_{i} \int d\b{V} f(\b{V}) P(\b{V} | \b{D}_{i},\b{\sigma}_i) \\
\Rightarrow\log {\cal L}\, &= \, \sum_{i} \log \int d\b{V} f(\b{V}) P(\b{V} | \b{D}_{i},\b{\sigma}_i) \\
&\equiv \,  \sum_{i} \log {\cal L}_i \label{eq:loglike} \, .
\end{align}
In calculating the likelihoods, ${\cal L}_i$, we assume a Schwarzschild distribution function, and normally distributed error in proper motion. The unknown radial velocity is integrated over analytically. Explicit expressions for ${\cal L}_i$ are given in appendix \ref{sec:likeeqns}.  

Again, the dispersion tensor and mean were constrained to reduce the number of parameters. We use flat priors on the dispersion and asymmetric drift. The expression for the likelihood was used to calculate the maximum likelihood estimate of the dispersion tensor, while errors were estimated from a MCMC using Metropolis-Hastings sampling. When the constraints on the dispersion tensor and mean velocity were relaxed this did not substantially alter the results, aside from the larger errors, particularly in the underpopulated bins due to the reduced degrees of freedom. In particular, the results are insensitive to allowing vertex deviation.

\begin{figure}
\includegraphics[width=\linewidth]{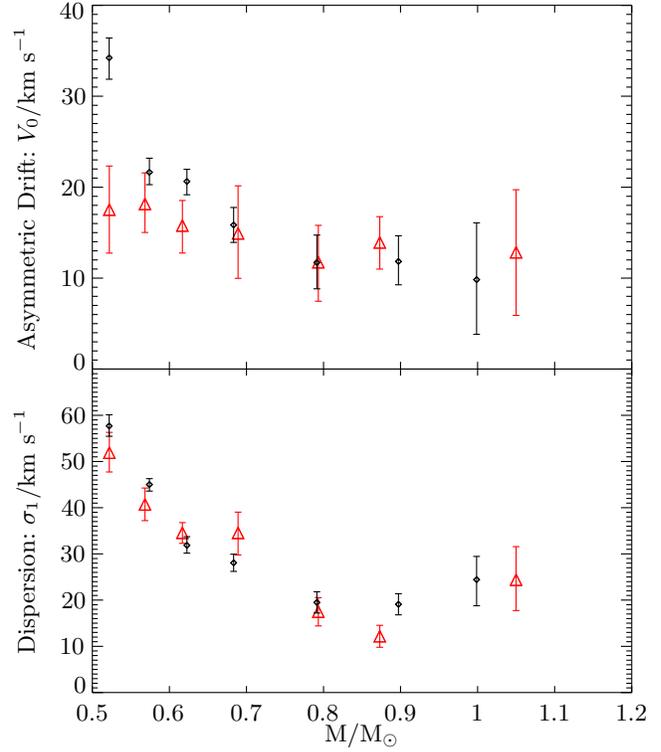}
\caption{Dispersion of SDSS (black) and PG (red) white dwarfs against mass calculated using the using the method described in section \ref{sec:mcmc}. Each bin is plotted at its mean mass.} 
\label{fig:results}
\end{figure}

\begin{figure}
\includegraphics[width=\linewidth]{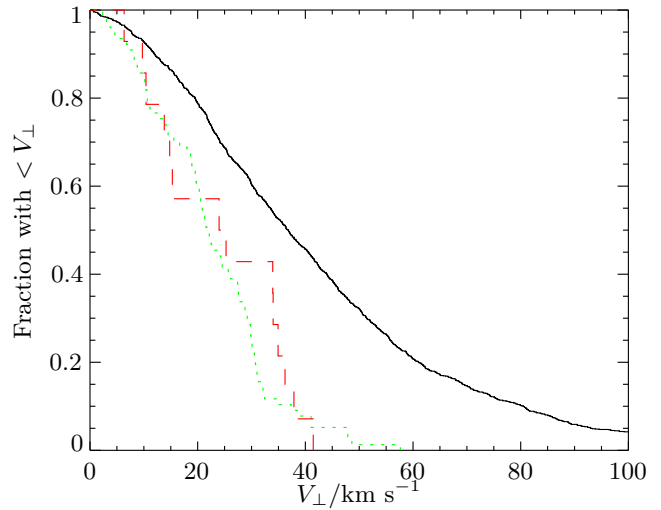}
\caption{Cumulative transverse velocity distribution of the combined SDSS and PG surveys. Low mass white dwarfs (\mrange{0.5}{0.75}, with $M=M_1+M_2$) as solid black,  high mass white dwarfs ($M>0.95\,\msun$) as dashed red, and intermediate mass white dwarfs (\mrange{0.75}{0.95}) as dotted green.} 
\label{fig:vperpgroup}
\end{figure}

The fitting results for the SDSS and PG samples are summarised in table \ref{tab:results} and plotted in figure \ref{fig:results}. In addition, in figure \ref{fig:vperpgroup} the raw transverse velocities measured from the proper motions for three groups of white dwarfs are shown. The lowest mass white dwarfs, \mlow{0.45}, are expected to be predominantly formed through binary evolution and have a binary white dwarf partner.
This potentially introduces errors into their photometric distances and so we do not consider them beyond simply stating the fitting results in table \ref{tab:results}. 

\begin{table*}
\caption[]{Kinematic fitting results from the PG and SDSS samples described in section \ref{sec:sample} using the methods of sections \ref{sec:dehnen} and \ref{sec:mcmc}\label{tab:results}. $M_{\rm low}$ and $M_{\rm high}$ are in units of $\msun$, while $\sigma_1$ and $V$ are in\kms. $N$ is the number of white dwarfs in each mass bin.}
\begin{tabular}{| c l | c c c c c c c | c c c c c c c|}
\hline
              &                &  \multicolumn{5}{c}{PG}                                                  &\multicolumn{5}{c}{SDSS} \\
$M_{\rm low}$ & $M_{\rm high}$ & $N$ & \multicolumn{2}{c}{\citet{Dehnen:1998}} & \multicolumn{2}{c}{MCMC} & $N$ & \multicolumn{2}{c}{\citet{Dehnen:1998}} & \multicolumn{2}{c}{MCMC} \\
              &                &     & $\sigma_1$ & $V$                        & $\sigma_1$ & $V$         &     & $\sigma_1$ & $V$                        & $\sigma_1$ & $V$         \\
\hline
$0.30$ & $0.40$ & $      5$ & $          47\pm          12$ & $          22\pm          13$ & $          48\pm          10$ & $          18\pm          12$ & $     70$ & $          53\pm           3$ & $          33\pm           4$ & $          40\pm           3$ & $          34\pm           4$\\
$0.40$ & $0.47$ & $     20$ & $          49\pm           6$ & $          27\pm           7$ & $          49\pm           7$ & $          28\pm           6$ & $     62$ & $          68\pm           4$ & $          38\pm           6$ & $          70\pm           6$ & $          38\pm           6$\\
$0.47$ & $0.55$ & $     35$ & $          47\pm           4$ & $          18\pm           6$ & $          51\pm           4$ & $          17\pm           4$ & $    333$ & $          56\pm           1$ & $          34\pm           2$ & $          57\pm           2$ & $          34\pm           2$\\
$0.55$ & $0.60$ & $     53$ & $          37\pm           2$ & $          17\pm           3$ & $          40\pm           3$ & $          18\pm           3$ & $    482$ & $          46\pm           1$ & $          20\pm           1$ & $          45\pm           1$ & $          21\pm           1$\\
$0.60$ & $0.65$ & $     51$ & $          37\pm           2$ & $          16\pm           3$ & $          34\pm           2$ & $          15\pm           2$ & $    239$ & $          33\pm           1$ & $          20\pm           1$ & $          31\pm           1$ & $          20\pm           1$\\
$0.65$ & $0.75$ & $     23$ & $          33\pm           3$ & $          15\pm           4$ & $          34\pm           4$ & $          14\pm           5$ & $     91$ & $          26\pm           1$ & $          16\pm           2$ & $          28\pm           1$ & $          15\pm           1$\\
$0.75$ & $0.85$ & $      9$ & $          16\pm           2$ & $          11\pm           4$ & $          17\pm           3$ & $          11\pm           4$ & $     30$ & $          16\pm           1$ & $          11\pm           2$ & $          19\pm           2$ & $          11\pm           2$\\
$0.85$ & $0.95$ & $     10$ & $          12\pm           2$ & $          15\pm           2$ & $          12\pm           2$ & $          13\pm           2$ & $     28$ & $          18\pm           1$ & $          12\pm           2$ & $          19\pm           2$ & $          11\pm           2$\\
$0.95$ & $1.44$ & $      5$ & $          22\pm           5$ & $          14\pm           7$ & $          24\pm           6$ & $          12\pm           6$ & $      9$ & $          19\pm           3$ & $           9\pm           5$ & $          24\pm           5$ & $           9\pm           6$\\
\hline
\end{tabular}

\end{table*}

%-----------------------------------------------------------------------
\section{Expectations from Single Star Evolution}
\label{sec:sse}
%-----------------------------------------------------------------------

%-----------------------------------------------------------------------
\subsection{Analytic}
\label{sec:sseanalytic}
%-----------------------------------------------------------------------

In this section we describe the reasons for the relationship WD mass and dispersion within a simple analytic model, before moving onto the more complex Monte Carlo simulations of section \ref{sec:montesse}.

Within the framework of single star evolution (SSE) an ensemble of white dwarfs with the same mass would be expected to have a dispersion $\sigma(t_{\rm TOT})$, where $\sigma(t)$ is the disk heating relation, and $t_{\rm TOT}$ is the total age of the white dwarf including its precursor lifetime (i.e. total pre-white dwarf stellar lifetime). Here $t_{\rm TOT}$ will be given by $t_{\rm TOT}=t_{\rm WD} + t_{\rm SSE}(M_{\rm i}(M_{\rm WD}))$ where $t_{\rm WD}$ is the cooling age of the white dwarf and $t_{\rm SSE}(M_{\rm i}(M_{\rm WD}))$ is the total precursor lifetime, which is a function of the white dwarf mass through the initial-final mass relation (IFMR) $M_i(M_f)$. Two components of this prediction are particularly uncertain: the disk heating relation and the IFMR. We discuss these now.

The best constraints on the IFMR come from open clusters. Spectroscopic fits of the masses of white dwarfs give the final mass. The initial mass is estimated using isochrone fitting to the main sequence turnoff to calculate the age of the cluster, which finally allows  the corresponding initial mass to be inferred using the precursor lifetime \citep{Catalan:08}. This method has succeeded in producing IFMRs with a typical uncertainty of less than 20\%. The strong dependance of the precursor lifetime on mass however makes this a considerable uncertainty in the dispersion relation.

The most accurate data on the disk heating relation is given in \citet{Nordstrom:2004} from an analysis of F and G dwarfs with radial velocities and Hipparcos data, although this data still permits a range of heating models \citep{Seabroke:2007}. However, for consistency, we instead use the disk heating models estimated in \citet{Just:2010}, since we also use their companion star formation histories.  

\begin{table*}
\begin{minipage}{180mm}
\renewcommand{\thempfootnote}{\alph{mpfootnote}}
\caption{Model input parameters for the models of single star evolution (SSE).}\label{tab:ssemod}
\centering
\begin{tabular}{lllll} 
\hline
Model & $\sigma(t)~/\kms$ & $M_{\rm i}(M_{\rm WD})~/~\msun$ & $t_{\rm SSE}(M_{\rm i})~/~{\rm Gyr}$ & SFR$(t)$\footnote{In units of ${\rm M}_\odot~{\rm pc}^{-2}~{\rm Gyr}^{-1}$. Not used in the analytic SSE simulation of section \ref{sec:sseanalytic}.}  \\
\hline
A & $ 66 \left(\frac{0.5+t/{\rm Gyr}}{0.5+12}\right)^{1/2}$ \footnote{\citet{Just:2010} model C\label{foot:justc}. Disk age $12\gyr$. \citet{Girardi:00} models use metal enrichment from \citet{Just:2010} model C.} & \multicolumn{2}{c}{From \cite{Hurley:2000}, solar metallicity.} & $3.25 ~~ ^{\ref{foot:justc}}$\\
B & $ 62 \left(\frac{0.32+t/{\rm Gyr}}{0.32+10}\right)^{1/2}$ \footnote{\citet{Just:2010} model D\label{foot:justd}. Disk age $10\gyr$. \citet{Girardi:00} models use metal enrichment from \citet{Just:2010} model D.} &  \multicolumn{2}{c}{From \cite{Hurley:2000}, solar metallicity.} & $7.68 \exp( -t / 8~{\rm Gyr}) ~~ ^{\ref{foot:justd}}$\\
C & $ 66 \left(\frac{0.5+t/{\rm Gyr}}{0.5+12}\right)^{1/2}~^{\ref{foot:justc}}$ & From \citet{Catalan:08} & From \citet{Girardi:00}$^{\ref{foot:justc}}$ & $3.25 ~~ ^{\ref{foot:justc}}$\\
D & $ 62 \left(\frac{0.32+t/{\rm Gyr}}{0.32+12}\right)^{1/2}~^{\ref{foot:justd}}$ & From \citet{Catalan:08} & From \citet{Girardi:00}$^{\ref{foot:justd}}$ & $7.68 \exp( -t / 8~{\rm Gyr}) ~~ ^{\ref{foot:justd}}$ \\
\hline
\end{tabular}
\end{minipage}
\end{table*}

\begin{figure}
\includegraphics[width=84mm]{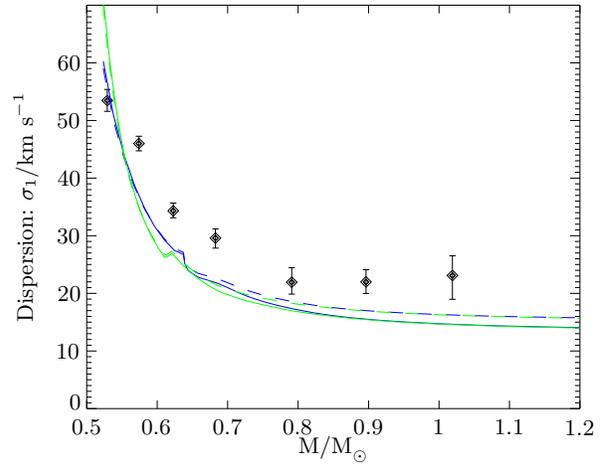}
\caption{Predicted dispersions from models A-D, as dashed blue, solid blue, dashed green and solid green respectively. Also plotted are SDSS white dwarfs (black error bars) with the data taken from the MCMC column of table \ref{tab:results}.} 
\label{fig:sseanal}
\end{figure}

The effect of these model uncertainties are shown in figure \ref{fig:sseanal} for the models described in table \ref{tab:ssemod}. Qualitatively the results appear to agree with the predicted relations: for white dwarfs more massive than $0.75\,\msun$ the white dwarf progenitors precusor lifetime is short and there is little dependance of the kinematics on mass. Below $0.75\,\msun$ the dispersion sharply increases as the progenitor lifetime approached $1\,$Gyr and longer where the disk heating is significant. 

However, while qualitatively the results in figure \ref{fig:sseanal} are consistent, there is quantitative disagreement. To assess this disagreement we turn to a more sophisticated Monte Carlo treatment.

%-----------------------------------------------------------------------
\subsection{Monte Carlo}
\label{sec:montesse}
%-----------------------------------------------------------------------
As a quantitative check of our results in section \ref{sec:sseanalytic} we have performed a Monte Carlo simulation of the production, kinematics, and observation of the white dwarfs in the solar neighbourhood, as described in this section. We also describe the simulated selection and observation of these white dwarfs by SDSS and PG. We perform this simulation to assuage fears that our results could be impacted by effects such as selection biases.

\begin{figure}
\centering
\includegraphics{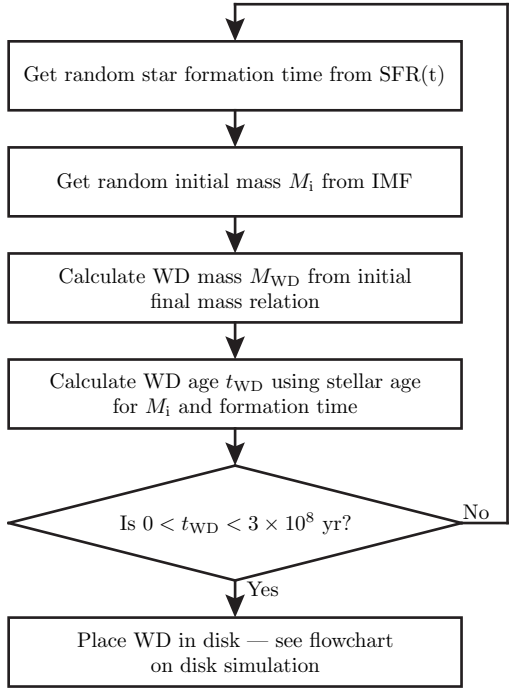}
\caption{Flowchart illustrating the process of simulating white dwarfs formed from single star evolution (SSE). If a star reaches the final stage, then it is placed in the disk using a process described by the flowchart shown in figure \ref{fig:diskflow}.} 
\label{fig:sseflow}
\vspace{6mm}
\end{figure}

\begin{figure}
\centering
\includegraphics{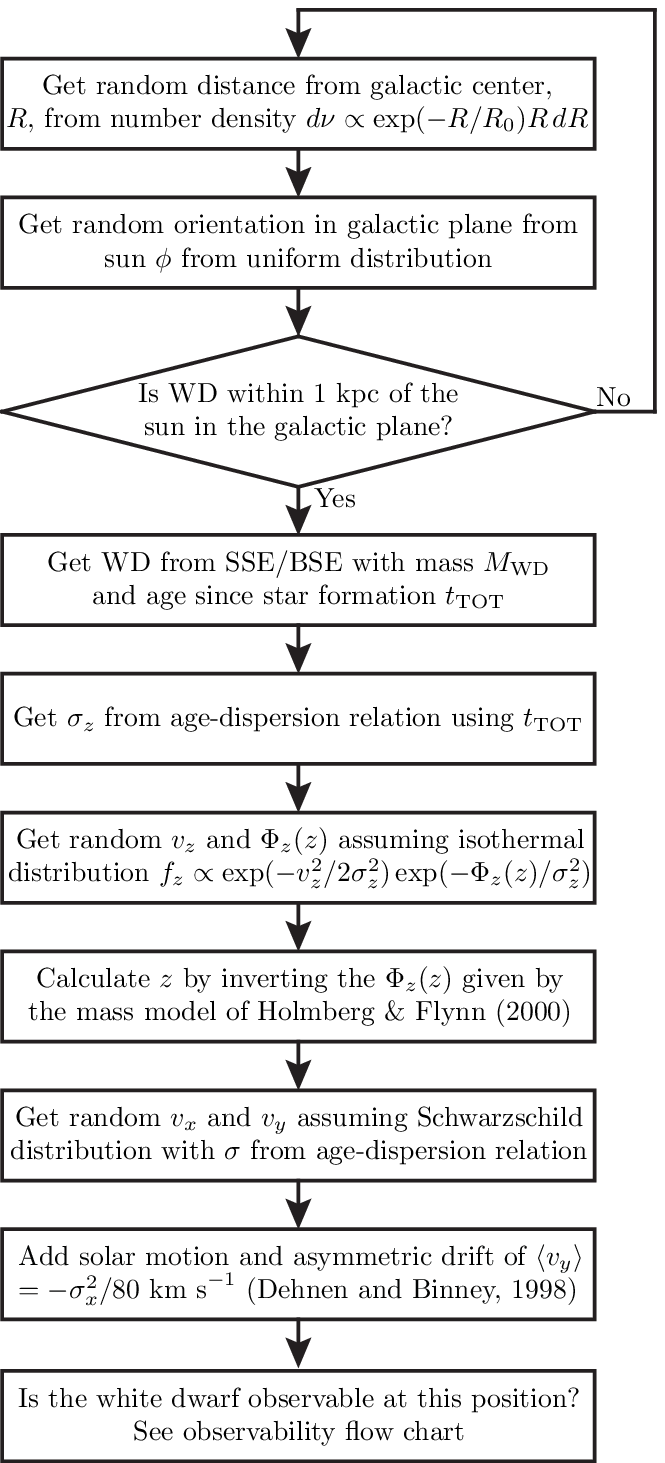}
\caption{Flowchart illustrating the process of placing white dwarfs in the galactic disk and picking their velocity. This process is undertaken if a star reaches the final stage of the flowchart shown in figure \ref{fig:diskflow}. If a star reaches the final stage of this flowchart, the observability is finally determined using the algorithm described in the flowchart shown in figure \ref{fig:obsflow}} 
\label{fig:diskflow}
\end{figure}

\begin{figure*}
\centering
\subfigure[SDSS.]{\label{fig:obsflowsdss}\includegraphics{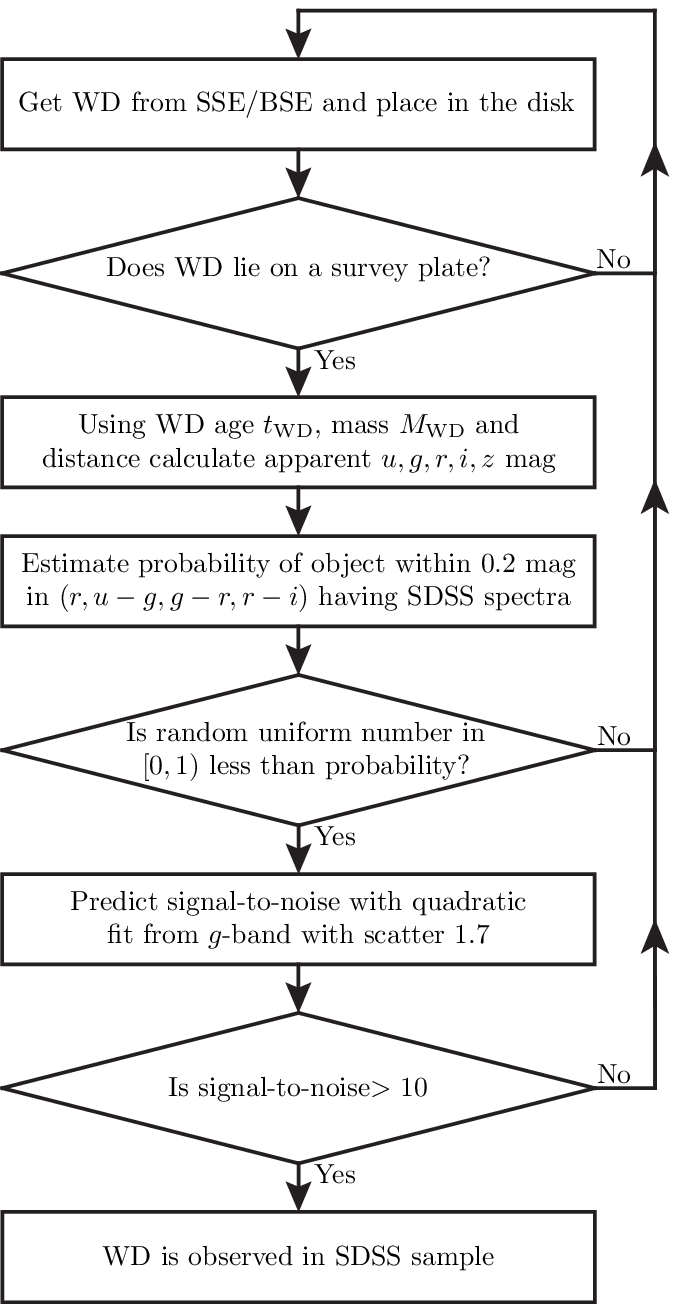}}
\quad
\subfigure[Palomar-Green (PG).]{\label{fig:obsflowpg}\includegraphics{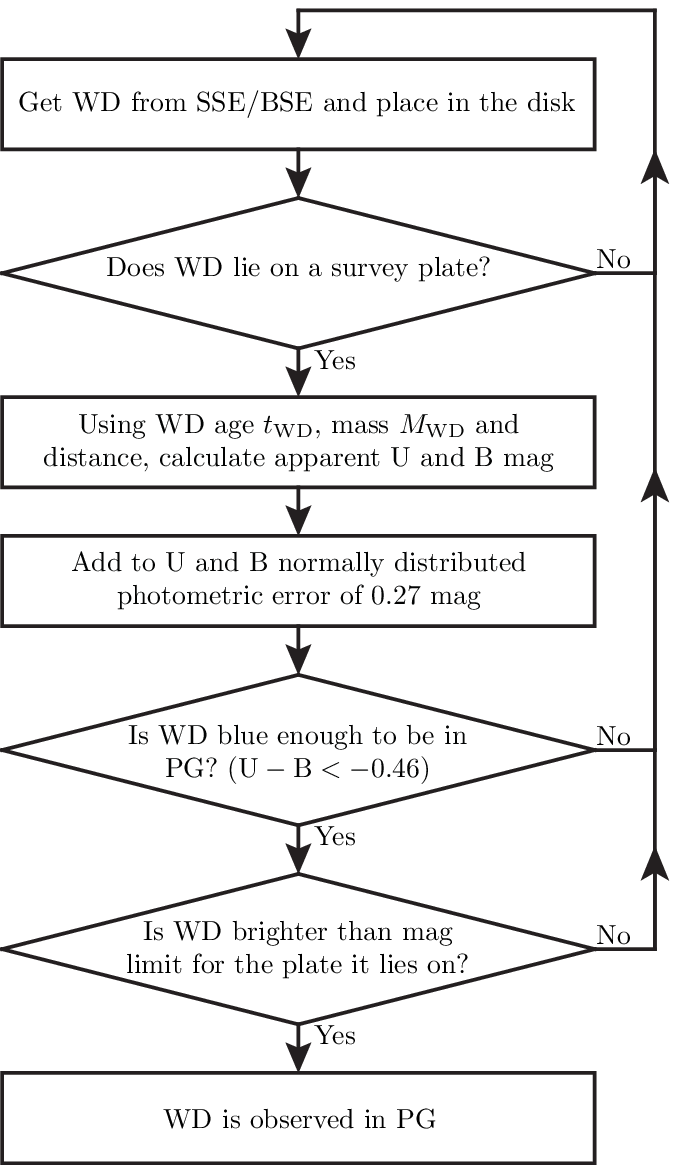}}
\caption{Flowchart illustrating the process of simulating whether white dwarfs are observed. This process is undertaken if a star has reached the final stage of the flowchart shown in figure \ref{fig:diskflow}. \label{fig:obsflow}} 
\end{figure*}

\begin{figure*}
\includegraphics[width=\linewidth]{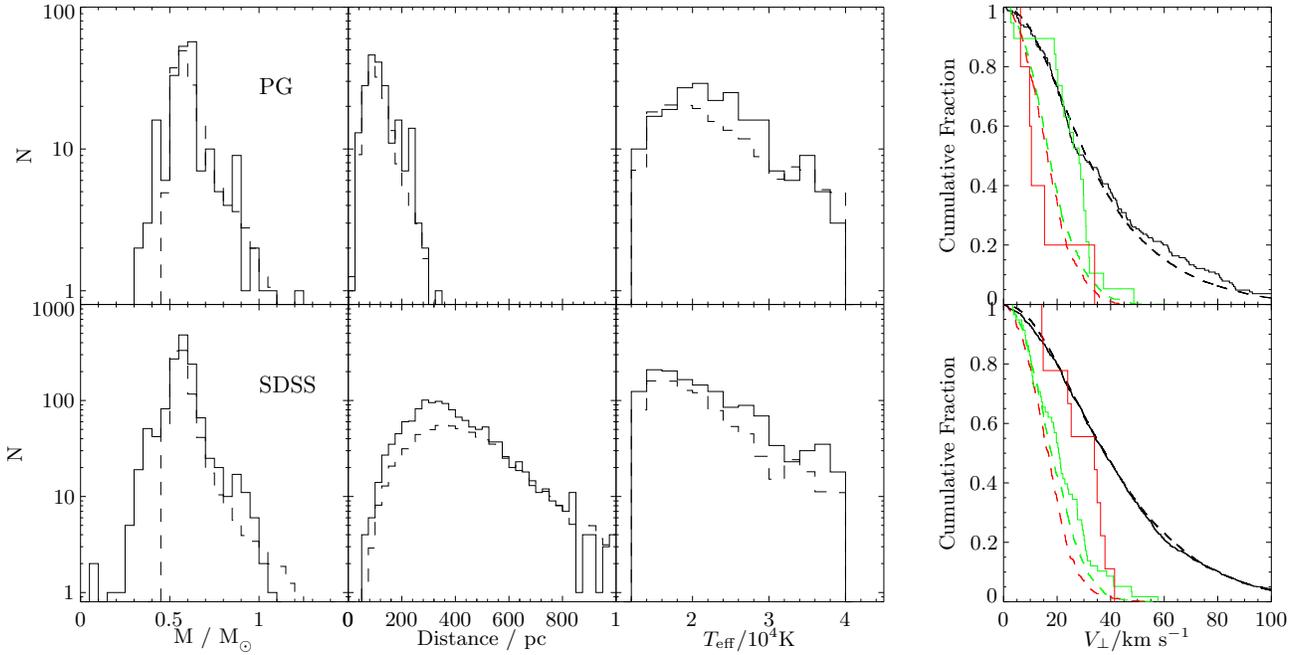}
\caption{Comparison between the observed white dwarfs as the solid lines, and the Monte Carlo simulations of single star evolution described in section \ref{sec:montesse} as the dashed lines. The upper panels show the PG survey and the lower panels the SDSS survey. The rightmost panel shows the cumulative transverse velocity distribution. In this panel colors are as in figure \ref{fig:vperpgroup}. Specifically low-mass white dwarfs (\mrange[M_1+M_2]{0.5}{0.75}) are plotted in black, high-mass ($M_1+M_2>0.95\,\msun$) white dwarfs in red, and intermediate mass white dwarfs (\mrange[M_1+M_2]{0.75}{0.95}) in green. The simulation plotted is model D from table \ref{tab:ssemod}.} 
\label{fig:ssemonte}
\end{figure*}

This process is somewhat involved, and so for clarity it is summarised in the flow charts in figures \ref{fig:sseflow}--\ref{fig:obsflow}. The final results of the Monte Carlo simulation are compared with the white dwarf sample in figure \ref{fig:ssemonte}.

{\sc Picking Stars:} The initial mass was drawn from a Kroupa IMF and one of two star formation histories (Table \ref{tab:ssemod}). If this resulted in a white dwarf at the present time with an age less than $3\times10^8$ years, and a temperature between 13,000K and 40,000K using the cooling models of \citet{Wood:1995} as explained in section \ref{sec:sample}, then it was included in the simulation. See figure \ref{fig:sseflow} for synopsis.

{\sc Placing Stars in Disk:} If a star has been included in the simulation, it is given a velocity dispersion taken from the previously described disk heating models of table \ref{tab:ssemod} and axis ratios of the velocity ellipsoid of 1:1/1.4:1/2.2 \citep{Dehnen:1998}. Its velocity in the disk was drawn from a Gaussian with these widths and it was placed in the plane of the Galaxy using a radial exponential disk with a scale length of 2.5\kpc. Since the furthest $>0.47\msun$ WD projected into plane is less than 1 kpc, only WDs placed within this distance are simulated further.

For an isothermal population the vertical position, $z$, and velocity, $v_z$, are given by
\begin{align}
f_z(E_z)& \propto \exp(-E_z/\sigma_z^2) \notag \\
&\propto\exp(-v_z^2/2 \sigma_z^2) \exp(-\Phi_z(z)/\sigma_z^2) \, ,
\end{align}
where $\Phi_z$ is the gravitational potential. Each star's velocity is thus drawn from a Gaussian with standard deviation given by the previously calculated $\sigma_z$, while $z$ is chosen by first drawing $\Phi_z(z)$ from an exponential distribution with scale $\sigma_z^2$, and then inverting this to calculate $z$. We use the mass models of \citet{Holmberg:2000} for $\Phi_z(z)$.

This process of placing white dwarfs in the local galactic disk is summarised in figure \ref{fig:diskflow}.

{\sc White Dwarf Observability:} As a result of this process, each white dwarf has an assigned galactic position and velocity, together with its mass and age. It is then assessed whether it is likely to be observed in either the SDSS or PG survey as follows: First its galactic position is translated to a right ascension, RA, and declination, dec, and, unless this falls on one of the PG plates or the SDSS DR4 spectroscopic plates, the probability of observation is zero. 

For white dwarfs in the PG survey the apparent U and B magnitude is calculated from the models of \citet{Holberg:2006} with a $0.27$ mag error added to each to mimic the photometric errors in PG \mbox{\citep{Liebert:2005}}. If it is bluer than $U-B=-0.46$ and brighter than the $B$ band magnitude limit for the PG plate on which it lies then it is considered observed. 

For SDSS the spectroscopic targeting is more complex \citep{Kleinman:2004}, and the strategy was to construct an empirical observational probability for a star at each magnitude and colour. A four dimensional table of probability of spectroscopic follow up was constructed in $(r,u-g,g-r,r-i)$ grouped in $0.2$ mag bins from the SDSS DR4 clean photometry.  The expected spectroscopic signal-to-noise was calculated using a quadratic least squares fit to the observed signal-to-noise ratio as a function of $g$-band magnitude together with normally distributed scatter in signal-to-noise with standard deviation of 1.7. If the signal-to-noise ratio was greater than 10 it was included in the mock sample.

Finally, measurement errors in mass of $0.03\,\msun$ and proper motion errors of 5.6\masyr are introduced. 

The process of assessing  if each white dwarf is observed by the PG or SDSS surveys is summarised in figure \ref{fig:obsflow}.  In all simulations we simulate a total of $\sim2\times10^{11}$ objects.

\begin{figure}
\includegraphics[width=\linewidth]{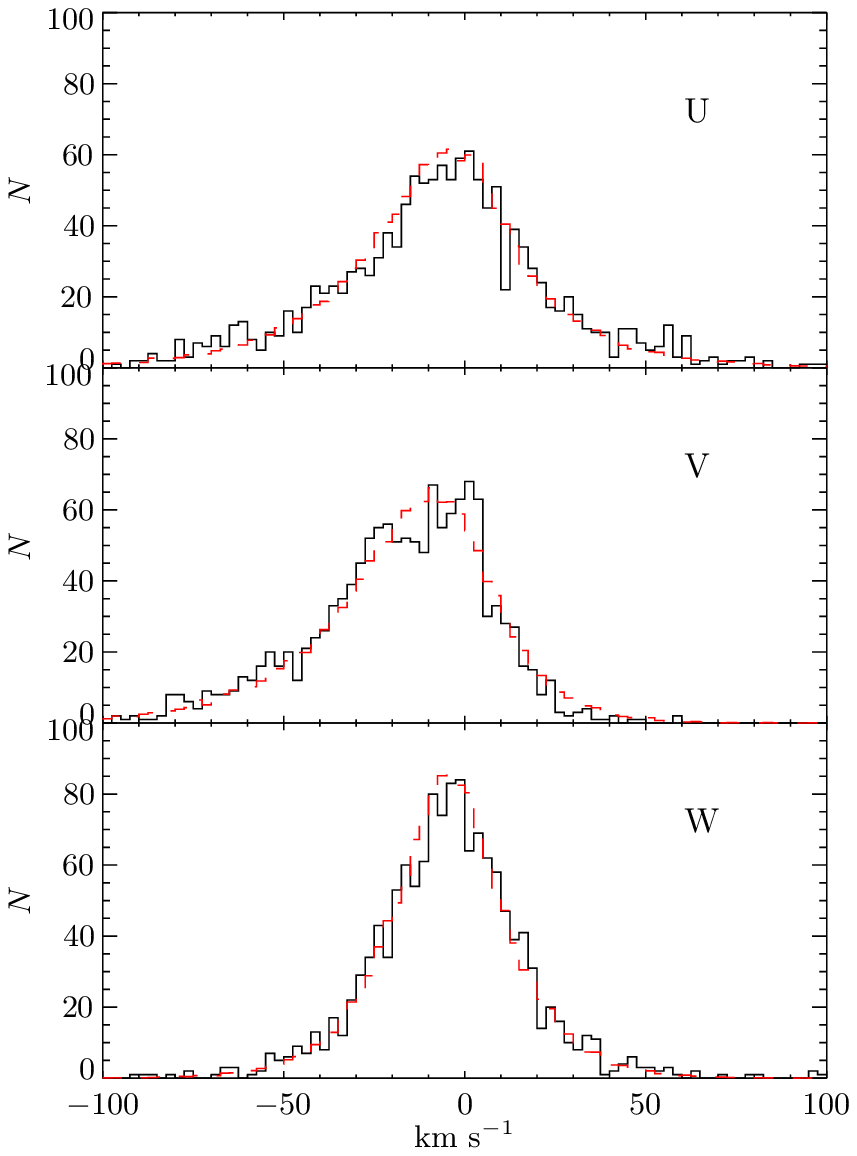}
\caption{Histograms showing the agreement between the observed and simulated velocity distribution in $U$, $V$, $W$ directions of SDSS WDs. The black line is the observed distribution, while the dashed red line is the distribution of the SSE simulation for model C. Zero radial velocity is artificially assumed, and number of simulated WDs is normalised to the number observed. $U$ is directed towards the galactic centre, $V$ in the direction of galactic rotation, and $W$ towards the north Galactic pole.}
\label{fig:uvwhist}
\end{figure}

{\sc Monte Carlo Results:} The results of this simulation are shown in figure \ref{fig:ssemonte}. As a further check that the simulated white dwarfs have the correct kinematics we plot the distributions in the $U$, $V$ and $W$ directions (directed towards the galactic centre, in the direction of galactic rotation, and towards the north Galactic pole respectively) in figure \ref{fig:uvwhist}.

The results of the singe star evolution (SSE) simulation, described in this section, closely agree with the observations, modulo the normalisation factor. We do not concern ourselves with this overall normalisation, however the normalisation factor is typically $\lesssim 2$. The simulation also does not produce white dwarfs below $\approx 0.47\,\msun$, which are generally expected to form through binary evolution. As may be expected from the analytic models plotted in figure \ref{fig:sseanal}, the models in table \ref{tab:ssemod} all produce white dwarfs that reasonably closely explain the observed samples and their kinematics and so we only plot the results of only one representative model in figure \ref{fig:ssemonte}.

%-----------------------------------------------------------------------
\section{Expectations from Binary Star Evolution}
\label{sec:bse}
%-----------------------------------------------------------------------

It has been suggested that the majority of high-mass white dwarfs were formed from mergers of binary white dwarfs, both on the basis of their number density \citep{Liebert:2005} and a possible peak at $~1\,\msun$ \citep{Vennes:1999}. To test this hypothesis we use two binary evolution codes (discussed in section \ref{sec:bec}) to perform binary population synthesis (described in section \ref{sec:bsepop}), and ultimately what fraction of the sample is likely to have had a binary WD progenitor (section \ref{sec:bsevssse}).

%-----------------------------------------------------------------------
\subsection{Binary Evolution Codes}
\label{sec:bec}
%-----------------------------------------------------------------------
To address the considerable uncertainties in binary evolution, two binary evolution codes were used. Specifically, the \bse code described in \citet{Hurley:2000}, and the \seba code described in \cite{Nelemans:2001}. Both codes use the same approach to modelling binary evolution: semi-analytic fits to the structure and evolution of isolated stars are combined with prescriptions for interactions between the stars. 

There are four key initial conditions that govern the evolution of a binary: the initial primary mass $M_{\rm 1i}$, the initial secondary mass $M_{\rm 2i}$ (or equivalently the mass ratio $q_{\rm i}=M_{\rm 1i}/M_{\rm 2i}$), the initial binary semi-major axis $a_{\rm i}$ and the initial eccentricity $e_{\rm i}$.

\begin{figure}
\includegraphics[width=84mm]{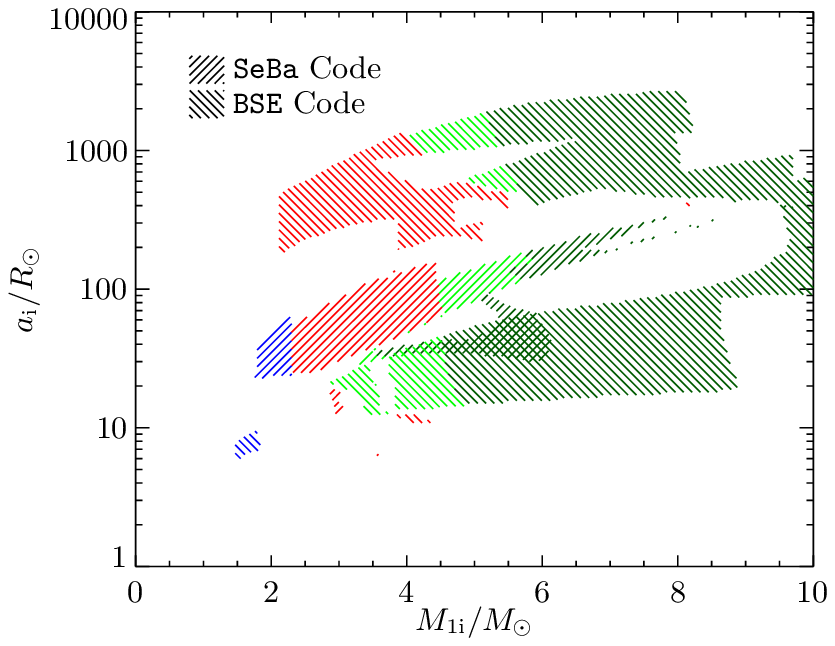}
\caption{Comparison of the WD+WD merger outcomes from the \seba and \bse codes with their default prescriptions for binary evolution. All simulations use an initial mass ratio of $q_{\rm i}=0.5$ and eccentricity of $e_{\rm i}=0$. Green corresponds to CO+CO, red He+CO and blue He+He. The lighter green are sub-Chandrasekhar ($M_1+M_2<1.4\,\msun$) mergers, and the dark green super-Chandrasekhar.} 
\label{fig:bse2dcomp}
\end{figure}

One slice through the four-dimensional space of initial conditions $(M_{\rm 1i}, q_{\rm i}, a_{\rm i}, e_{\rm i})$ showing those conditions which result in the merger of a pair of white dwarfs is shown in figure \ref{fig:bse2dcomp}. 

The differences between the \bse code and \seba code in figure \ref{fig:bse2dcomp} are striking, and are largely due to the different binary evolution prescriptions, and in particular the treatment of the Roche lobe overflow (RLOF) and common envelope (CE) phases. 

For the specifics of the treatment of the RLOF phase and its treatment in the \bse and \seba codes we refer the reader to \citet{Hurley:2000} and \cite{Nelemans:2001} respectively.
There is also considerable uncertainty in the treatment of the important CE evolution phase. The most fundamental difference between the codes is the treatment of the first phase of mass transfer. \bse uses the most commonly used prescription for common envelope evolution known as the $\alpha$ parameterisation, for both phases of mass transfer. \seba however, by default uses the $\gamma$ parameterization. We refer the reader to  \cite{Nelemans:2001} for the details of these parameterisations.

To assess the result of the considerable uncertainties in binary evolution on the merger time distribution, and therefore the resultant velocity distributions,  we have used four models across the two binary evolution codes.
These models are summarised in table \ref{tab:bsemodels}.

\begin{table}
\centering
\begin{tabular}{c c c c c}
\hline
Model & Evolution Code & CE Prescription & $\alpha_{\rm CE} \lambda$ & $\gamma$ \\
\hline
i  & \bse  & $\alpha \alpha$ & 2 & -   \\
ii & \bse  & $\alpha \alpha$ & 1 & -   \\
iii& \seba & $\gamma \alpha$ & 2 & 1.5 \\
iv & \seba & $\alpha \alpha$ & 2 & -   \\
\hline
\end{tabular}
\caption{Summary of the four binary evolution models considered. The \bse code is that described by \citet{Hurley:2000}, and the \seba code is described in \citet{Nelemans:2001}. The common envelope (CE) prescription describes how the two phases of common envelope evolution are treated. For example $\gamma\alpha$ describes treatment of the first phase through the $\gamma$ parameterisation and the second through the $\alpha$ parameterisation. We refer the reader to \citet{Nelemans:2001} for the definition and descriptions of these parameterisations.}
\label{tab:bsemodels}
\end{table}

%-----------------------------------------------------------------------
\subsection{Binary Population Synthesis}
\label{sec:bsepop}
%-----------------------------------------------------------------------

We now describe our method of binary population synthesis.

We use the same distributions in the parameters $(M_{\rm 1i}, q_{\rm i}, a_{\rm i}, e_{\rm i})$ as \citet{Han:1998} and \citet{Nelemans:2001} with the exception of the IMF for which we use a \citet{Kroupa:2001} IMF as opposed to a \citet{Miller:1979} IMF. 
For reference the probability distributions are:
\begin{alignat}{3}
P(M_{\rm 1i}) &\propto M_{\rm 1i}^{-1.35} \qquad&&  0.8 < M_{\rm 1i} \leq 10 \notag \, ,\\
P(q_{\rm i}) &\propto {\rm const.} && 0<q\leq1 \, ,\label{eq:bseprobs} \\
P(\log a_{\rm i}) &\propto {\rm const.} && 0<\log a_{\rm i}/\rsun\leq5 \notag  \, , \\
P(e_{\rm i}) &\propto e_{\rm i} && 0\leq e_{\rm i}<1 \notag \, .
\end{alignat}

Our approach to simulating the results of binary star evolution is to first produce a 4-dimensional grid of binary simulations in the parameters $(M_{\rm 1i}, q_{\rm i}, a_{\rm i}, e_{\rm i})$. Grid points were linearly spaced in $M_{\rm 1i}$ between 0.8 and $10\,\msun$, linearly spaced in $q_{\rm i}$ between 0 and 1, logarithmically spaced in $a_{\rm i}$ between 1 and $10^4\rsun$, and linearly spaced in $e_{\rm i}^2$ between 0 and 1. The grid size used was a $25\times25\times50\times10$ grid in $(M_{\rm 1i}, q_{\rm i}, a_{\rm i}, e_{\rm i})$, respectively. With this choice of grid combined with the distributions in equation \ref{eq:bseprobs} the population synthesis is particularly simple: an initial primary mass is drawn from the \citet{Kroupa:2001} IMF and a random binary from the closest corresponding $(q_{\rm i}, a_{\rm i}, e_{\rm i})$ slice is chosen. In all simulations a total of $\sim 10^{13}$ objects are places in the disk.

\begin{figure}
\centering
%\psfrag{A}[c][c]{equation \ref{eq:bseprobs}}
\includegraphics{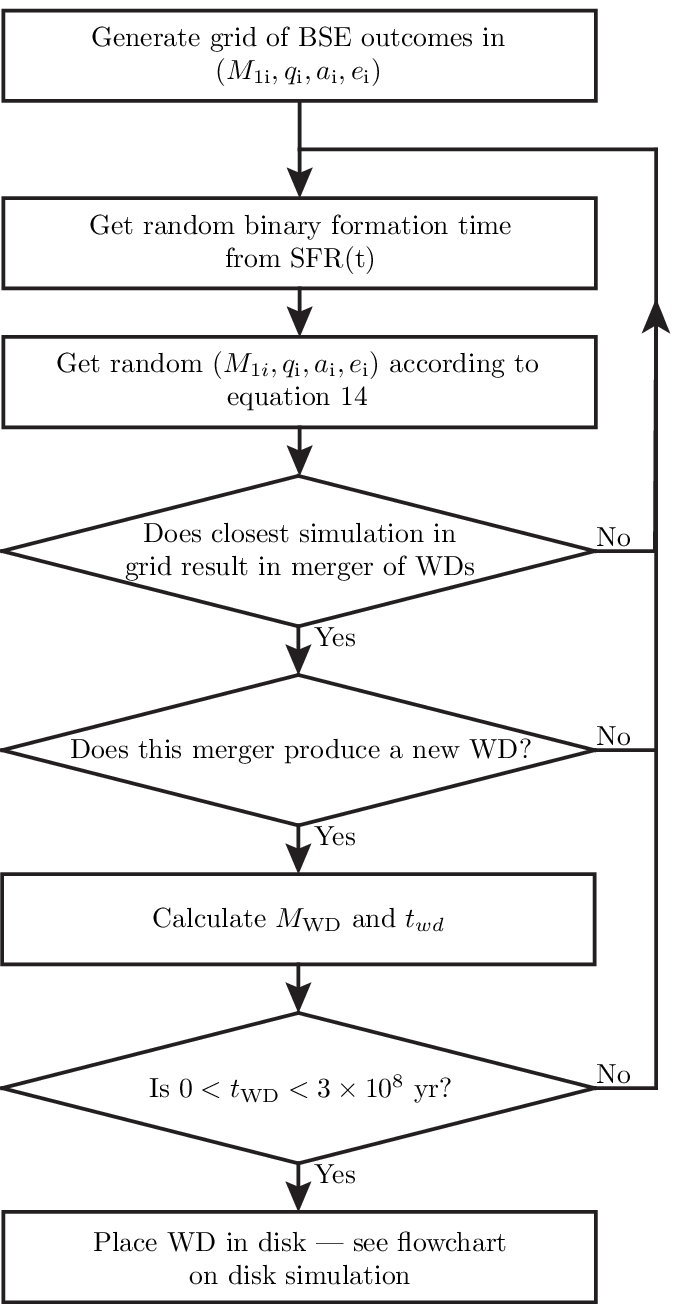}
\caption{Flowchart illustrating the process of simulating white dwarfs formed from binary star evolution.} 
\label{fig:bseflow}
\end{figure}

The process of simulating stars formed from binary evolution is summarised in figure \ref{fig:bseflow}. 

\begin{figure}
\centering
\includegraphics[width=\linewidth]{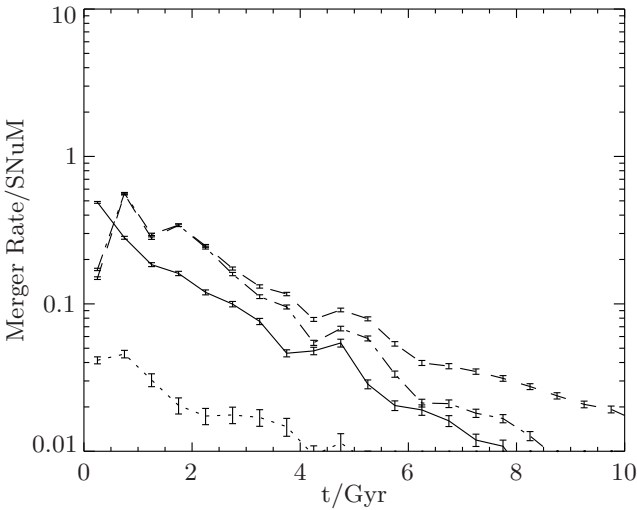}
\caption{Merger rates of CO+CO WDs with sub-Chandrasekhar total mass following a burst of star formation. The error bars are purely statistical due to the finite size of the simulated binary grid. Lines are models i-iv as solid line, dotted line, dashed line, and dash-dot line respectively. The models are described in table \ref{tab:bsemodels}. ${\rm SNuM}\equiv 1/({\rm 100~yr})~/(10^{10}~\msunm)$. }
\label{fig:cocomergerrate}
\label{fig:coco}
\end{figure}

In what follows we concern ourselves with the merger of CO+CO white dwarfs, since these are the mergers proposed to result in $\gtrsim1\,\msun$ white dwarfs. Thus, in figure \ref{fig:cocomergerrate} we plot the rate at which pairs of white dwarfs with sub-Chandrasekhar total mass merge as calculated from our binary population synthesis of the four models in table \ref{tab:bsemodels}. Note that the overall normalisation can be very different. In particular, model ii uses a relatively efficient CE prescription with $\alpha_{\rm CE} \lambda=1$ for both phases of mass transfer. This in turn results in a smaller range of initial separations that will ultimately result in a gravitational radiation driven WD merger. Despite the differences in overall rate between the models they all display a similar {\em distribution} of merger times. This is because, apart from at early times, the merger time is dominated by the time to merge by gravitational radiation. This is a strong function of separation, $a$, specifically $t_{\rm GW} \propto a^4$. As a result, at late times, the merging WDs originally formed a narrow range in separation at WD+WD birth. Approximating this as a power law, $\frac{dN}{da}\propto a^{\epsilon}$ leads to a merger rate $\frac{dN}{dt}=\frac{dN}{da}\frac{da}{dt}\propto t^{-(3-\epsilon)/4}$, and so for a wide range of $\epsilon$ the merger rate declines as $\frac{dN}{dt} \sim t^{-1}$ \citep{Maoz:2010}.

%-----------------------------------------------------------------------
\subsection{Proportion of high-mass White Dwarfs Formed in Mergers}
\label{sec:bsevssse}
%-----------------------------------------------------------------------

To assess the possible proportion of high-mass white dwarfs that formed through mergers, the CO+CO merger products with  \mrange[M_1+M_2]{0.95}{1.4} from the binary population synthesis, are subjected to the same process as the single population synthesis results {\it i.e.} they are placed locally in the disk according to the method summarised in figure \ref{fig:diskflow} and their observability in the SDSS and PG samples assessed according to figure \ref{fig:obsflow}.

\begin{figure}
\includegraphics[width=84mm]{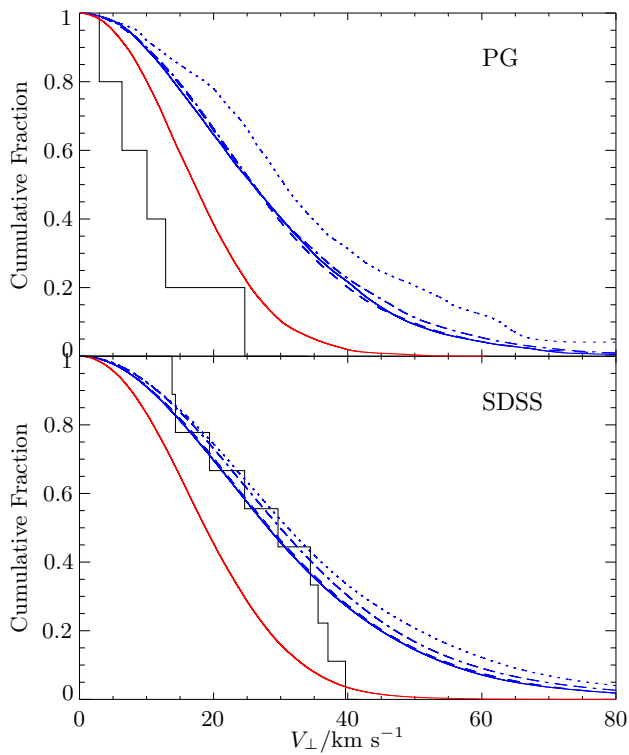}
\caption{Predicted distribution of transverse velocities observed in SDSS and PG resulting from the merger of CO+CO WDs with $0.95<M_1+M_2/\msun1.4$. Lines are the \bse code with $\alpha_{\rm CE}  \lambda=2$ (solid blue line, model i), the \bse code with $\alpha_{\rm CE} \lambda=1$ (dotted blue line, model ii) the \seba code using the $\gamma\alpha$ common envelope prescription (dashed blue line, model iii) and the \seba code using the $\alpha\alpha$ prescription (dash-dot blue line, model iv). Both \seba models use $\alpha_{\rm CE} \lambda=2$ and $\gamma=1.5$. The red line is the predicted distribution of transverse velocities resulting from single star evolution to a $0.95<M/\msun<1.4$ white dwarf according to model A in table \ref{tab:ssemod}, and the black line are the observed distributions. All BSE models use a constant SFR and the disk heating relation of model A in table \ref{tab:ssemod}.}
\label{fig:bsevt}
\end{figure}

We assume that no mass is ejected during the merger so that resultant white dwarf has mass $M_{\rm WD}=M_1+M_2$. We also assume that the merger reheats the white dwarf sufficiently that the white dwarf has a cooling age of
\[
t_{\rm WD}=t_{\rm form}-t_{\rm merge}
\] 
where $t_{\rm form}$ is the time prior to the present at which the binary initially formed, and $t_{\rm merge}$ is the length of time it took for the merger to occur, including the precursor lifetime. The resulting cumulative transverse velocity of \mrange[M_1+M_2]{0.95}{1.4} CO+CO merger products are shown in figure \ref{fig:bsevt}. 

In figure \ref{fig:bsevt} and the following we have combined the PG and SDSS samples to improve the statistics. We combine the Monte Carlo results by the empirical proportions of WDs in this sample i.e. the observed PG to SDSS ratio of 5:9. Note however there is a possible discrepancy between the two samples in this high mass bin. In particular the SDSS sample has few low velocity ($<14$\kms) white dwarfs (see the bottom right panel of figure \ref{fig:ssemonte}), and this results in a 12\% probability that they are drawn from the same distribution.

The distribution of transverse velocities in figure \ref{fig:bsevt} shows that despite the uncertainties in binary evolution resulting in very different binary histories (figure \ref{fig:bse2dcomp}) and overall merger rates (figure \ref{fig:cocomergerrate}), the resultant velocity distributions are very similar. This is a result of the $\sim t^{-1}$ merger time distribution at late times discussed previously. 

The results in figure \ref{fig:bsevt} naturally lead the question of what fraction of mergers is consistent with the data to be addressed. We wish to assess the fraction of high-mass galactic white dwarfs formed by binary mergers (BSE) which we parameterise by $\theta$. This results in a fraction $1-\theta$ from single star evolution (SSE).
To assess a value of $\theta$ for a given SSE and BSE Monte Carlo realisation we first calculate the galactic formation rate of high-mass WDs from SSE and BSE in this realisation, which we denote $\ssefr$ and $\bsefr$, respectively. Then, for both PG and SDSS we make $\alpha$ copies of the BSE objects simulated as observed, and $\beta$ copies of objects simulated as observed from SSE. Assuming that equal numbers of objects were simulated in both the BSE and SSE realisations, then the two simulated samples combined have a galactic BSE fraction of
\[
\theta = \frac{\beta \bsefr}{\beta \bsefr + \alpha \ssefr} \;.
\]
To test whether the data is consistent with this realisation, we use the two sample Anderson-Darling statistic \citep{Pettitt:1976}. The Anderson-Darling test considers the difference between the samples across the entire distribution, and so is more statistically powerful that the more commonly used Kolmogorov-Smirnov test which depends only on the extremum. The number of simulated white dwarfs is always much larger, by at least a factor of ten, than the number observed.  

\begin{figure*}
\subfigure[The left column shows the cumulative distribution of transverse velocities of high-mass (\mlim{0.95}) white dwarfs in the SDSS and PG survey. The dashed-dot lines are the predictions of SSE model C and the dashed lines are the predictions of BSE model iv. The right column shows, for each fractional galactic formation fraction from BSE, $\theta$, the probability that the velocity distribution is consistent with the data using the Anderson-Darling statistic for the PG sample, $P_{\rm PG}$, and the SDSS sample, $P_{\rm SDSS}$. The fiducial $\theta$ is the fiducial predicted galactic fraction from BSE model iii compared to SSE model C with 50 per cent binary fraction.]{\includegraphics[width=0.475\linewidth]{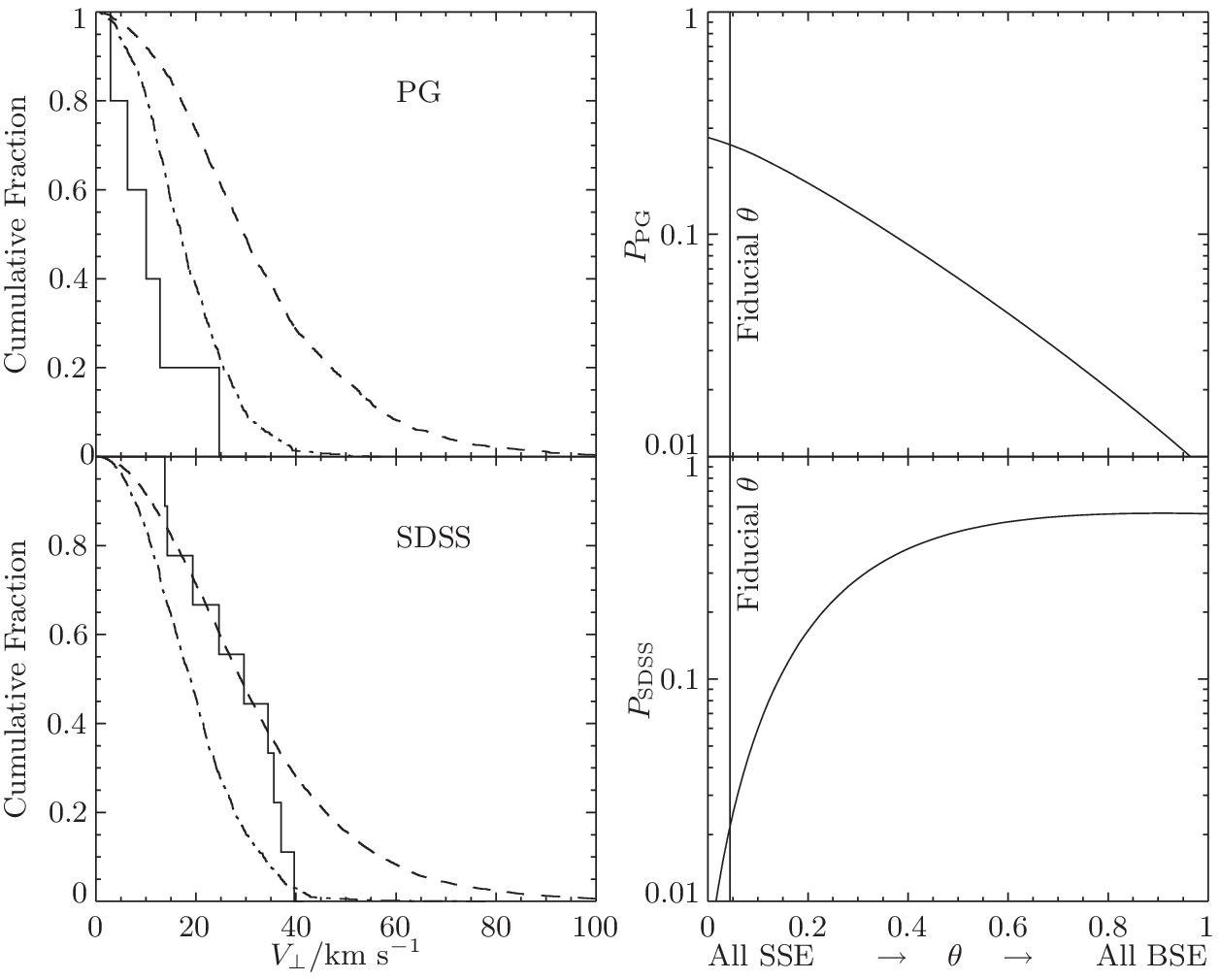}\label{fig:bsematrix}}
\hspace{0.025\linewidth}
\subfigure[The combined probability that each value $\theta$ is consistant with both the PG and SDSS samples. Calculated by the product of the probabilities in the right column of figure \ref{fig:bsematrix}.]{\includegraphics[width=0.475\linewidth]{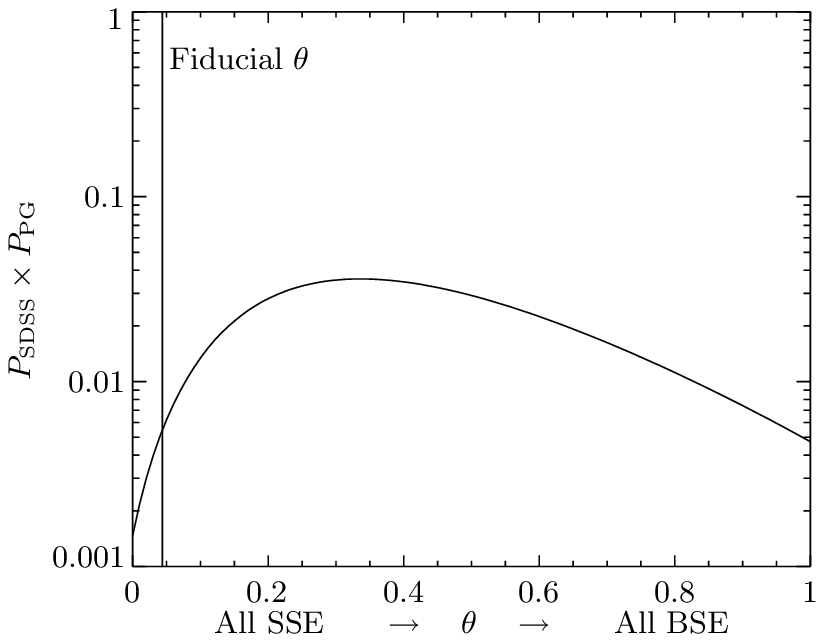}\label{fig:bsecomb}}
\caption{Plots showing the calculation of the galactic formation fraction of high-mass white dwarfs formed in mergers during binary star evolution in model C compared to single star evolution model iii. }
\end{figure*}

The results for one particular choice of SSE and BSE model are shown in figure \ref{fig:bsematrix}. In figure \ref{fig:bsecomb} we show the combined probability that the PG and SDSS samples are consistent with each value of $\theta$. In table \ref{tab:bseresults} we summarise the results of this procedure for the range of the BSE and SSE models described in tables \ref{tab:ssemod} and \ref{tab:bsemodels}.

The results in table \ref{tab:bseresults} show that for the majority of models the fiducial numbers of white dwarfs formed via SSE and BSE are consistent with the data. The results taken at face value would also appear to show that, for most models, at the 1 per cent probability level, high-mass white dwarfs must come from a combination of single star evolution and mergers of high-mass white dwarfs. This appears artificial however: From the right column of figure \ref{fig:bsematrix} the PG sample is consistent with all SSE, while the SDSS sample that has a low probability of arising purely from SSE.

This is a result of the lack of low velocity ($<14$\kms) white dwarfs in the SDSS sample. It may be that the lack of low velocity white dwarfs in SDSS is a statistical anomaly, since the number of objects is small. In theory this would be taken account of in the analysis described above, however young stellar objects can display prominent substructure in their kinematics as a result of moving groups \citep[e.g.][]{Dehnen:1998b}. This would have the result of both reducing the effective sample size, and producing a very different velocity distribution than the Schwarzschild distribution assumed in the SSE Monte Carlo. There are indications that this is the case, since when the SDSS objects are plotted in the $U-V$ plane (assuming zero radial velocity) 7 of the 9 objects lie in the negative U, negative V quadrant. Depending on the unobserved radial velocity, many of these could have kinematics consistent with the Pleiades and Hyades moving groups. Indeed it has been shown that the the white dwarf GD 50, has a velocity and cooling age consistent with a Pleiades origin \citep{Dobbie:06}.

That the data rules out a white dwarf merger origin for the majority of high-mass white dwarfs appears more secure, despite the apparent consistency of the SDSS sample with the BSE simulations: The PG sample is entirely consistent with SSE, and neither sample contains a high-mass white dwarf travelling at $>50$\kms~ which would be convincing evidence of a BSE origin for some high mass white dwarfs. This is not surprising, since the expected number of merger products observed in PG and SDSS ($N_{\rmn{BSE}}$ in table \ref{tab:bseresults}) is significantly smaller than the observed number of objects. 

We note that a simpler empirical test for the origin of the high-mass white dwarfs is suggested by figure \ref{fig:vperpgroup}. The distribution of high-mass white dwarfs is consistent with the velocity distribution of the intermediate group that displays the kinematics of young objects at the 13 per cent level by the Anderson-Darling test. This ignores the selection effects which the Monte Carlo simulation addresses, but does suggest that the entire combined group of high mass white dwarfs is broadly consistent with SSE.
	
\begin{table*}
\begin{tabular}{c c c c c c c c c c c c}
\hline
\multirow{2}*{SFR} & \multirow{2}*{SSE Model}  & \multirow{2}*{BSE Model} & \multirow{2}*{$\Gamma_{\rm BSE}$} & \multirow{2}*{$\Gamma_{\rm SSE}$} & \multirow{2}*{$\theta_{\rm fid}$} & \multicolumn{2}{c}{PG} & \multicolumn{2}{c}{SDSS} & \multirow{2}*{$P(\theta_{\rm fid})$} & \multirow{2}*{$\theta(P>0.01)$}\\
& & & & & & $N_{\rm SSE}$ & $N_{\rm BSE}$&$N_{\rm SSE}$ & $N_{\rm BSE}$& \\
\hline
\multirow{4}*{Const} & C & i 
&    0.0006&      0.03&      0.02&           7&       0.1&          17&       0.9&      0.02&0.09-0.9\\
& C & ii 
&    0.0001&      0.03&     0.005&           7&      0.02&          17&       0.1&     0.005&0.08-0.8\\
& C & iii 
&     0.001&      0.03&      0.04&           7&       0.3&          17&        2.&      0.04&0.08-0.8\\
& C & iv 
&     0.001&      0.03&      0.04&           7&       0.3&          17&        2.&      0.04&0.09-1.\\
\hline
\multirow{4}*{Exp} & D & i 
&    0.0007&      0.02&      0.04&           6&       0.1&          13&        1.&      0.04&0.1-0.9\\
& D & ii 
&    0.0002&      0.02&      0.01&           6&      0.02&          13&       0.2&      0.01&0.2-0.8\\
& D & iii 
&     0.001&      0.02&      0.07&           6&       0.3&          13&        3.&      0.07&0.1-0.9\\
& D &  iv 
&     0.001&      0.02&      0.06&           6&       0.3&          13&        2.&      0.06&0.2-0.9\\
\hline
\end{tabular}
\caption[]{Summary of the results of calculation of the fraction of high-mass white dwarfs formed in mergers compares to single star evolution. The SSE models are described in table \ref{tab:ssemod} and the BSE models are described in table \ref{tab:bsemodels}. $\Gamma_{\rm BSE}$ is the galactic formation rate (in ${\rm yr}^{-1}$) from binary star evolution assuming that the merger of two CO white dwarfs with combined mass between $0.95\,\msun$ and $1.4\,\msun$ results in a high-mass white dwarf. $\Gamma_{\rm SSE}$ is the galactic formation rate from single star evolution. $\theta$ is the galactic fraction of high-mass white dwarfs formed from BSE so that the fiducial value is given by $\theta_{\rm fid}\equiv\frac{\Gamma_{\rm SSE}}{\Gamma_{\rm BSE}+\Gamma_{\rm SSE}}$. The numbers $N_{\rm SSE}$ and $N_{\rm BSE}$ are the predicted observed numbers from SSE and BSE evolution respectively in the PG and SDSS samples. $P(\theta_{\rm fid})$ is the probability that {\em both} the PG and SDSS velocity distributions are consistent with $\theta_{\rm fid}$ using the Anderson-Darling statistic. $\theta(P>0.05)$ is the range of $\theta$ values which have a probability of being consistent with the data greater than 1 per cent. The fiducial value of $\theta$ is calculated assuming a 50 per cent binary fraction (i.e., two-thirds of all stars formed in binaries). Both SSE and BSE models use the same disk heating model and star formation history: model C of table \ref{tab:ssemod} for the constant SFR models, and model D for the exponential. \label{tab:bseresults}}
\end{table*}

%------------------------------------------
\section{Scale Heights}
\label{sec:scaleh}
%------------------------------------------

One of the key results of this study is that hot white dwarfs of mass $\gtrsim 0.75\,\msun$ had much shorter main sequence lifetimes than their lower mass counterparts, and hence their kinematics are characteristic of young stars. A direct result of this is that these higher mass white dwarfs will have reduced scale height. This is vitally important to consider when calculating the formation rate as a function of mass using local samples such as in \mbox{\cite{Liebert:2005}} or \cite{Kepler:2007} or producing galactic white dwarf simulations such as \cite{Nelemans:2001}. 

Unfortunately, neither the SDSS or PG sample allow accurate direct determination of the scale height of each white dwarf population, particularly the rare and less luminous high-mass groups. Instead, here we list the expected scale height by comparison with the SSE models that appear to accurately describe the kinematics. We do this to allow simple initial corrections without resorting to the simulations of the type performed in this work. The scale height, $h$, was defined through
\begin{equation}
\nu(z)=\nu_0\,\text{sech}^2 \left(\frac{z}{2h}\right) \, ,
\label{eq:scaleh}
\end{equation}
where $\nu(z)$ is the stellar number density in terms of the height above the plane of the galactic disk, $z$. The scale height, $h$, was estimated by constraining equation \ref{eq:scaleh} to give both the correct overall number and central WD density, $\nu_0$. We choose this method since the most common usage of the scale height is to calculate galactic birthrates from local densities. The results are give in table \ref{tab:scaleh}. Note that the higher mass groups smaller scale height results in a local density enhanced by more than a factor of two over the more common low-mass group. In particular, the apparent excess of high-mass white dwarfs found in the PG survey \citep[discussed in section 6 of][]{Liebert:2005} can be naturally explained by their lower scale height, which causes a high abundance in this relatively local survey. That the number of high-mass white dwarfs is consistent with single star expectations in PG is confirmed by the number of expected white dwarfs from single star evolution in table \ref{tab:bseresults}.

\begin{table}
\caption[]{Scale heights, $h$, defined through equation \ref{eq:scaleh} for three different mass groups. $h$ is calculated by matching the central density and overall number to the simulations described in section \ref{sec:montesse}.}\label{tab:scaleh}
\centering
\begin{tabular}{c c c}
\hline
$M_{\rm low}~/\msunm$ & $M_{\rm high}~/\msunm$ & $h~/~{\rm pc}$ \\
\hline
 0.45& 0.75&         120\\
 0.75& 0.95&          58\\
 0.95& 1.40&          54\\
\hline
\end{tabular}
\end{table}

%-----------------------------------------------------------------------
\section{Summary}
\label{sec:summary}
%-----------------------------------------------------------------------

We have analysed the kinematics of young ($<3\times 10^8~{\rm years}$) DA white dwarfs from both the PG and SDSS surveys and find a strong connection between their mass and kinematics: low-mass white dwarfs (\mrange[M_1+M_2]{0.45}{0.75}) display the kinematics of old stars, with higher velocity dispersion ($\sim 46\kms$) and asymmetric drift, while higher mass white dwarfs (\mrange[M_1+M_2]{0.75}{0.95}) display the kinematics of young stars with a velocity dispersion of only $\sim 19\kms$. We have shown in section \ref{sec:sse} that this is expected due to the shorter precursor lifetime of the more massive progenitors, and that there is agreement both on simple analytic grounds (section \ref{sec:sseanalytic}) and more quantitive Monte Carlo simulations of the PG and SDSS samples (section \ref{sec:montesse}).

A further key conclusion is that the white dwarf scale height and its variation with age and mass is vitally important to consider when calculating birth rates based on local samples (section \ref{sec:scaleh}).

In addition, we have separately analysed the highest mass white dwarfs (\mlim{0.95}, section \ref{sec:bse}), since it has been suggested that many of these formed as a result of the merger of two lower mass CO white dwarfs. We find at present a discrepancy in the SDSS velocity distribution where no high-mass white dwarfs with transverse velocity less than 14\kms~is detected. This results in a velocity distribution that within our statistical framework is inconsistent with purely single star evolution. We argue this is likely to an anomaly, either be a statistical, or a result of a number of these white dwarfs being members of moving groups. We find  that, even under the most optimistic binary evolution models, we would only expect to find $3$ white dwarfs formed via white dwarf binary mergers and that the apparent excess of high mass white dwarfs found in PG is caused by their reduced scale height. In addition, we note the kinematic `smoking gun' of some fraction of high-mass white dwarfs coming from binary evolution would be high-mass white dwarfs traveling at $>50$\kms, of which none are found in PG or SDSS. 

%-----------------------------------------------------------------------
\section{Acknowledgements}
%-----------------------------------------------------------------------

CW gratefully acknowledges many useful discussions with Nate Bode.

Support for this work was provided by NASA BEFS grant NNX-07AH06G. 

%-----------------------------------------------------------------------
\appendix
\section{Likelihoods}
\label{sec:likeeqns}
%-----------------------------------------------------------------------

Here we give our expressions for the proper motion likelihoods of an individual object. These largely follow  \mbox{\citet{Ratnatunga:1989}}, modified to include errors in proper motion. We ignore errors in sky position ($\ell$, $b$), which are small.

Assuming a Schwarzschild distribution function, then, in coordinates aligned with the principle axes of the velocity ellipsoid,
\be
	f(\b{V}) = \frac{1}{\sqrt{8 \pi^3} \sigma_1 \sigma_2 \sigma_3} \exp \left( -(\b{V} - \b{V_0})^{\rm T} \cdot \b{\Gamma} \cdot (\b{V} - \b{V_0}) \right) \ ,
\ee
where $\b{\Gamma}=\mathrm{diag}(1/2\sigma_1,1/2\sigma_2,1/2\sigma_3)$ and $\b{V_0}$ is the mean velocity. Ignoring errors in distance, we then rotate to axes aligned with the sky plane, and integrate over the unobserved radial velocity, which, in this case, is a nuisance parameter. 

We define, $\b{\Lambda}$, to be the dispersion tensor rotated into the coordinate system, $(\ell,b,d)$, aligned with the sky plane. This will be given by $\b{\Lambda}=\b{R}\cdot\b{\Gamma}$, where $\b{R}$ is a rotation matrix \citep[given explicitly as equation A4 in][]{Ratnatunga:1989}.  The  probability distribution, after integrating over the radial velocity as a nuisance parameter, is an ellipsoid in the sky plane
\ba
	p(v_l,v_b) = C' \exp \big[ &-\alpha (v_\ell - \bar{v_\ell})^2 - \beta (v_b - \bar{v_b})^2  \nonumber \\
	&   - 2 \gamma (v_\ell - \bar{v_\ell}) (v_b - \bar{v_b}) \big]\,, \label{eq:probvel}
\ea
where $\bar{v_\ell}$ and $\bar{v_\ell}$ are the components of $\b{V_0}$ in the directions of $l$ and $b$ (which can be obtained via $(\bar{v_{\ell}},\bar{v_b},\bar{v_d})=\b{R}\cdot\b{V_0}$) and $\alpha$, $\beta$, $\gamma$, and $C'$ are given by
\ba
\alpha &= \Lambda_{22} - \Lambda_{12}^2/\Lambda_{11}\,,\\
\beta &= \Lambda_{33} - \Lambda_{13}^2/\Lambda_{11}\,,\\
\gamma &= \Lambda_{23} - \Lambda_{12}\Lambda_{13}/\Lambda_{11} \,, \\
C'&=\sqrt{\alpha \beta - \gamma^2}/\pi \, .
\ea

For each object we have measurements of $v_l$ and $v_b$, together with an associated velocity error $\sigma$. Integrating over the `true' $v_l$ and $v_b$ gives the log likelihood used in equation \ref{eq:loglike} as 
\ba
	\log {\cal L}_i (v_\ell^{\rm obs},v_b^{\rm obs}) \equiv & \log \int d\b{V} f(\b{V}) P(\b{V} | v_\ell^{\rm obs},v_b^{\rm obs},\sigma) \nonumber \\
	 & = \log C''  - \frac{\delta}{(\alpha+\delta)(\beta+\delta) - \gamma^2} \times \nonumber \\
	\big[ &(\Delta v_b^2 + \Delta v_\ell^2) ( \alpha \beta - \gamma^2) + \nonumber \\ 
 	 & \delta ( \beta \Delta v_b^2 + \alpha \Delta v_\ell^2 + 2 \gamma \Delta v_\ell \Delta v_b )\big]\,, \label{eq:loglikefull}
\ea
where 
\ba
\delta &= 1/2\sigma^2 \,, \\ 
\Delta v_\ell&=v_\ell^{\rm obs} - \bar{v_\ell} \, , \\
\Delta v_b&=v_b^{\rm obs} - \bar{v_b} \, , \\
C'' &= C' \frac{\delta}{\sqrt{\pi}\sqrt{(\alpha+\delta)(\beta+\delta) - \gamma^2}} \\
&= \delta \sqrt{ \frac{\alpha \beta - \gamma^2}{\pi^3 [(\alpha+\delta)(\beta+\delta) - \gamma^2]}}  \, . 
\ea
Note that for small error, $\delta\rightarrow \infty$, and equation \ref{eq:loglikefull} reduces to the log of equation \ref{eq:probvel} as expected.

%-----------------------------------------------------------------------
\bibliographystyle{mn2e}

\bsp
\label{lastpage}
\end{document}